\documentclass[10pt]{article}
\usepackage{amsfonts}
\usepackage{amsmath}
\usepackage{amssymb}
\usepackage{graphicx}

\def \be {\begin{equation}}
\def \ee {\end{equation}}
\def \bea {\begin{eqnarray}}
\def \eea {\end{eqnarray}}
\def \nn {\nonumber}

\def \rr {\raise.35ex\hbox{\small $\prime$}\kern-.17em{\mbox{\large $\imath$}}}
\def \del {\partial}
\def \dels {\partial\kern-.5em / \kern.5em}
\def \As {{A\kern-.5em / \kern.5em}}
\def \Ds {D\kern-.7em / \kern.5em}

\def \a {\alpha}

\def \b {\beta}

\def \d {\delta}

\def \lam {\lambda}

\def \s {\sigma}

\def \om {\omega}

\def \th {\theta}

\def \t {\tau}

\def \sgn {\mbox{\small sgn}}

\begin{document}
 \begin{titlepage}

\begin{center}
\vskip .5in

\textbf{\LARGE Nonlocal Particles as Strings}

\vskip .5in
{\large Tai-Chung Cheng$^{1}$, 
Pei-Ming Ho$^2${}$^\dagger$, Tze-Kei Lee$^2$}
\vskip 15pt

{\small ${}^1$
Taipei Municipal Zhung-Lun High School,
Taipei, Taiwan, R.O.C. \\
\vskip .1in
\small ${}^2$
Department of Physics and Center for Theoretical Sciences, \\
National Taiwan University,
Taipei, Taiwan, R.O.C.}

\vskip .2in
\sffamily{
$^\dagger$pmho@phys.ntu.edu.tw}

\vspace{60pt}
\end{center}

\begin{abstract}

We find nonlocal particle theories with two dimensional conformal symmetry, 
including examples equivalent to the bosonic open string and closed string. 
This work provides a new approach to construct solvable consistent backgrounds in string theory. 

\end{abstract}
\end{titlepage}
\setcounter{footnote}{0}
\section{Introduction}

Many of the remarkable features of string theory 
are often attributed to the dimensionality of the string, 
since these features are not remotely shared by any known particle theory. 
For example, 
a necessity of the T-duality is the existence of winding modes, 
which is apparently absent in particle theory. 
Although it is also possible to take the viewpoint that string field theory 
is just a theory of infinitely many fields of particles, 
the spectrum of the fields should have an origin in a two dimensional conformal field theory. 
The conformal symmetry is needed to ensure a consistent perturbation theory 
free of UV divergence and ghosts. 
\footnote{
Nevertheless, 
UV-finite, unitary field theories with infinitely many particles 
can also be constructed 
without reference to the conformal symmetry \cite{HoTian}. 
}
It is generally believed that the two dimensional conformal symmetry 
is at the heart of many miracles of string theory. 

The main idea behind this work is that 
the distinction between strings and particles is blurred 
when we consider particle actions that are nonlocal. 
Usually we assume that the equation of motion for a bosonic degree of freedom 
is a second order differential equation, 
so that the phase space of a $D$ dimensional particle is $2D$. 
In contrast, the string, viewed as a collection of infinitely many points, 
has a phase space of infinite dimensions. 
But for an equation of motion involving the $n$-th derivatives, 
the phase space is $nD$ dimensional. 
When $n \rightarrow \infty$, 
it is possible that the phase space of a particle can be identified with that of a string.
It was shown in \cite{Ho} that the reparametrization symmetry 
of a particle's worldline approaches to the 2 dimensional conformal symmetry 
in the $n \rightarrow \infty$ limit. 
This strongly suggests the equivalence between 
some nonlocal particle theories and string theories. 

On the other hand, higher derivative theories have been known 
to suffer from many unphysical problems, 
such as non-unitarity, spectra unbounded from below, 
acausality etc. \cite{EW}. 
These problems are always present in any theory with 
higher derivatives of a finite order. 
Theories with infinite derivatives are hard to classify, 
and these problems may or may not be present. 
The higher derivative interactions in an effective theory 
should be treated in a perturbative approach \cite{ChengHoYeh}. 
Higher derivative interaction treated as a fundamental theory 
is unusual to most physicists.

In this paper, we find explicit (quadratic) nonlocal actions of particles 
which respect the two dimensional conformal symmetry (Sec. \ref{Conformal}). 
The conformal symmetry is realized as the reparametrization symmetry 
of the particle worldline. 
Noether's theorem guarantees that the symmetry generators 
obey the Virasoro algebra (Sec. \ref{Generators}). 

We give many examples to illustrate the basic formulation, 
and suggest how a new dimension naturally arises 
on the worldline (to turn it into a worldsheet) in many cases. 
A generic nonlocal particle theory may not be unitary 
but it is easy to, 
and we give explicit rules for how to,
contruct examples preserving unitarity (Sec. \ref{Unitarity}).
Several unitary nonlocal particle theories are discussed in some detail.
First we give a nonlocal particle theory which is equivalent to 
an open string in the trivial background (Sec. \ref{OpenString}).
This equivalence guarantees that this nonlocal theory is well defined.
\footnote{The nonlocal particle theory for worldsheet ghosts is also found 
(Sec. \ref{Ghosts}).} 
Next we give examples equivalent to open strings in nontrivial backgrounds 
(Sec. \ref{ThirdExample}).
These nontrivial backgrounds
can be described in terms of 
nonlocal interactions in the string worldsheet theory. 
Finally we define a nonlocal particle theory equivalent to a closed string 
in the flat background (Sec. \ref{ClosedString}).

Our study reveals a new approach to construct consistent, 
solvable backgrounds in string theory. 
In general these new backgrounds correspond to higher derivative interactions 
on the string worldsheet, 
and are usually evaded in the traditional worldsheet approach to string theory. 
Our study also shed some light on the conceptual question
whether our universe is made of particles or strings.

\section{Nonlocal actions with conformal symmetry} \label{Conformal}

\subsection{Canonical quantization of nonlocal particle action}

In order to have solvable equations of motion, 
we restrict ourselves to quadratic actions
\be \label{Sxy}
S = - \int dt \; x(t) f(\del_t) y(t). 
\ee
This action involves two scalar fields $x$ and $y$. 
The case of a single scalar field can be easily derived from this case 
by identifying $x(t)$ with $y(t)$. 

Let us carry out the canonical quantization. 
The variation of this action is 
\be \label{deltaSgeneral} 
\d S = \int dt \; \left\{ 
\d x [f(\del_t) y(t)] + \d y(t) [f(-\del_t) x(t)] 
+ \left( x(t) [ f(\del_t) \d y(t) ] - [ f(-\del_t) x(t) ] \d y(t) \right) \right\}, 
\ee
where the last term in $(\cdots)$ is a total derivative. 
The equations of motion are 
\be
f(-\del_t) x(t) = 0, \qquad f(\del_t) y(t) = 0. 
\ee
The most general solution of $x$ and $y$ are 
\be
x(t) = \sum_a x_a e^{-i k_a t}, \qquad
y(t) = \sum_a y_a e^{i k_a t} ,
\ee
where $\{k_a\}$ is the set of zeros of $f$: 
\be \label{Z}
{\cal Z} = \{ k_a \} = \{ k \, | \, f(ik) = 0 \}. 
\ee

The total derivative in (\ref{deltaSgeneral}) is
\be
x(t) [ f(\del_t) \d y(t) ] - [ f(-\del_t) x(t) ] \d y(t) = \frac{d}{dt} \theta, 
\ee
where
\be \label{theta}
\theta = \bullet \left( \frac{1\otimes f(\del_t) - f(-\del_t)\otimes 1}{1\otimes \del_t + \del_t \otimes 1} \right)
(x(t) \otimes \d y(t)). 
\ee 
Here $\bullet$ is the symbol for multiplication, i.e. $\bullet(f \otimes g) = fg$. 
The quotient of the functions of derivative on the right hand side of (\ref{theta}) 
is well defined as long as $f(\del_t)$ admits a Taylor expansion 
\be
f(\del_t) = \sum_{ n = 0 }^{\infty} f_n \del_t^n.
\ee
The sympletic two-form $\omega = \d \theta$ is then
\bea
\omega &=& \bullet \left( \frac{1\otimes f(\del_t) - f(-\del_t)\otimes 1}{1\otimes \del_t + \del_t \otimes 1} \right) 
(x(t) \otimes \d y(t)) 
\label{omega2} \\
&=& \sum_{n = 0}^{\infty} f_n \left\{ \bullet
\sum_{k = 1}^n \left[ (-\del_t)^{k-1} \otimes \del_t^{n-k} \right] (\d x(t) \otimes \d y(t)) 
\right\}. 
\eea

Assuming that all zeros of $f$ are simple (non-degenerate), 
we find
\be
\omega = \sum_a \dot{f}(ik_a) \d x_a \d y_a. 
\ee
To derive this expression we note that (\ref{omega2}) contains the factor 
\be
\bullet \left( \frac{1\otimes f(i k_b) - f(i k_a)\otimes 1}{1\otimes i k_b - i k_a \otimes 1} \right)
\ee
when acting on the term $x_a e^{-i k_a t} \otimes y_b e^{i k_b t}$ in $x \otimes y$. 
Since $f(i k_a) = f(i k_b) = 0$, this factor vanishes unless $k_a = k_b$. 
When $k_a = k_b$, it is obvious that 
this factor is supposed to be understood as the derivative of $f$, i.e., $\dot{f}(i k_a)$. 
From the viewpoint of both the symplectic two-form and the Hamiltonian, 
a pair of variables $(x_a, y_a)$ is independent of another pair $(x_b, y_b)$ unless $k_a = k_b$. 

If there are double poles at $k_a$, the general solution is of the form
\be
x = (x_a + p_a t) e^{-i k_a t} + \cdots, \qquad
y = (y_a + q_a t) e^{i k_a t} + \cdots, 
\ee
and the relevant part of the sympletic two-form will be 
\be \label{omega3}
\omega = \frac{1}{2} \ddot{f}(i k_a) (\d x_a \d q_a + \d y_a \d p_a) + \cdots .
\ee
This is easy to derive from (\ref{omega2}). 
One can consider this as the limiting case of two nearby simples zeros $k_a, k_b$ 
with the difference approaching to zero, $(k_a - k_b) \rightarrow 0$. 
Similarly, for triple poles or poles of higher order degeneracy, 
it is straightforward to derive the symplectic two-form from (\ref{omega2}) in the same way, 
although the theory would then have either a Hamiltonian unbounded from below 
or negative norm states.

\subsection{Particle actions with 2D conformal symmetry} \label{Generators}

There are nonlocal particle actions with conformal symmetry, 
which is realized as the symmetry of worldline reparametrization 
\be
x(t) \rightarrow x'(t) = x(t+\delta t). 
\ee
For reparametrizations in the harmonic basis
\be
\d t = \epsilon e^{int} + c.c., \qquad \forall n \in \mathbb{Z}, 
\ee
where $\epsilon \in \mathbb{C}$ 
and $c.c$ means complex conjugate,
the infinitesimal transformation of $x$ 
\be \label{transf}
\delta x(t) \equiv x'(t) - x(t) = \epsilon e^{-int} \dot{x}(t) + c.c., 
\ee
is generated by the differential operator 
\be
V_n = e^{-int} \del_t 
\ee
satisfying the classical Virasoro algebra 
\be
[V_m, V_n] = (m-n) V_{m+n}. 
\ee
In the above we have restricted the values of the index $n$ 
to be integers so that the algebra generated by $V_n$'s 
is no more than the Virasoro algebra. 
For $t$ compactified on a circle 
(this is of course not physical), 
we can always normalize $t$ so that $t\in[0, 2\pi)$, 
then $\{e^{int}\}_{n\in\mathbb{Z}}$ constitute 
a complete basis for functions of $t$ 
and the Virasoro algebra is equivalent to 
the full reparametrization symmetry of $t$. 
For the uncompactified time direction $t\in\mathbb{R}$, 
if we demand the most general reparametrization symmetry, 
$n$ can be any real number and 
the particle dynamics would have to be trivial. 
By restricting the values of $n$ to integers, 
we can still have many nontrivial models as shown below. 
Note that a length scale is introduced here 
corresponding to the shift $\Delta t$ in $t$ that will 
make all transformations in (\ref{transf}) trivial. 
We are using the unit system so that 
this length scale is normalized to $2\pi$. 

With these Virasoro generators imposed as constraints, 
the theory of the nonlocal particle has the gauge symmetry of conformal group, 
which is the defining feature of string theory. 

The symmetry (\ref{transf}) implies that if $x(t) = e^{-i z t}$ is a solution to the equation of motion, 
then $x(t) = e^{-i (z+n) t}$ must also be a solution for any integer $n$. 
\footnote{
There is an exception when $z_i \in \mathbb{Z}$. 
Due to the derivative of $x$ in the transformation rule 
(\ref{transf}), 
$k = 0$ does not have to be a zero of $f$.
We will give examples on this in Sec. \ref{3rdclass}.
}
As long as $f(i k)$ satisfies the following property: 
\be
\mbox{If} \quad f(i k_a) = 0, \quad \mbox{then} \quad f(i(k_a + n)) = 0 \quad \mbox{for all} \quad n \in \mathbb{Z},
\ee
the action (\ref{Sxy}) has the conformal symmetry (\ref{transf}). 
The zeros of the function $f$ must come in sequences 
\be \label{Z}
{\cal Z} 
= \{ z_i + n; \;n \in \mathbb{Z} \}. 
\ee
The zeros $z_i$ constitute a minimal set of zeros such that 
(\ref{Z}) includes all the zeros of $f$. 

The general solution of the equations of motion is
\be
x(t) = \sum_{k\in\cal Z} x_k e^{-i k t} 
, \qquad 
y(t) = \sum_{k\in\cal Z} y_k e^{i k t} 
\ee
Without loss of generality we can restrict $z_i$'s to lie within the range $[0, 1)$. 
If $f$ is symmetric, $f(-\del_t) = f(\del_t)$, 
$z_i$ always comes in pairs $(z_i, - z_i)$ except $z_i = 0$ or $1/2$. 
The symplectic two-form (\ref{omega3}) is now
\be \label{omega4}
\omega = 
\sum_{k\in\cal Z} \dot{f}(ik) \, \d x_k \d y_k .
\ee
Here and below we assume that all zeros of $f$ are non-degenerate and real. 

The conserved charges are $\theta$ 
with $\d x$, $\d y$ given by the symmetry transformation (\ref{transf}):
\be \label{Qxy}
Q_n = 
\sum_{k\in\cal Z} 
i (k + n) \dot{f}(i k) x_{k} y_{(k+n)}.
\ee

According to (\ref{omega4}), we define the creation, annihilation operators 
\be \label{abxy}
a_{-k} = (\dot{f}(ik)k)^{1/2} \, x_{k}, \qquad 
b_k = (\dot{f}(ik)k)^{1/2} \, y_{k}, 
\ee
so that the symeplectic two-form is equivolent to  
\be \label{omegaab}
\omega = 
\sum_{k \in \cal Z} 
\frac{\d a_{-k} \d b_k}{k}.
\ee
If $k = (z_i + n) > 0$, $a_{-k}$ is the creation operator and
$b_k$ is the annihilation operator. 
If $k = (z_i + n) < 0$, $a_{-k}$ is the annihilation operator and 
$b_k$ is the creation operator. 
In order for (\ref{abxy}) to make sense, 
we assume here that 
\be \label{fdotk}
\dot{f}(ik) k > 0
\ee
at all zeros of $f$ because this is the condition 
that the Hamiltonian is bounded from below.
A sequence of zeros $\{z_i + n\}$ is viewed as {\em space-like} 
if (\ref{fdotk}) holds.
If the opposite holds
\be
\dot{f}(ik) k < 0
\ee
at a sequence of zeros, 
that sequence should be viewed as time-like, 
and the definition of $a_k$, $b_k$ in (\ref{abxy}) 
should be modified by adding factors of $(-1)$
in the square roots.
The contribution of a time-like sequence of zeros of $f$ to
the Hamiltonian is bounded from above.
For the theory to be consistent, 
we should demand that the Virasoro constraints
be sufficient to eliminate unwanted states 
so that the spectrum of physical states is bounded from below.
No-ghost theorem still needs to be proved.

The conserved charges are
\be \label{QB}
Q_n = \sum_{k\in\cal Z} B(n, k) \, a_{-k} b_{k+n}, 
\ee
where 
\be \label{B}
B(n, k) = 
\left(\frac{\dot{f}(ik)}{\dot{f}(i(k+n))}\frac{k+n}{k}\right)^{1/2}. 
\ee
In particular, the Hamiltonian is 
\be
Q_0 = \sum_{k \in \cal Z} a_{-k} b_k, 
\ee
which has a non-negative spectrum upon quantization after normal ordering.
Using the property of $B(n, m)$ 
\be \label{BBB}
B(n, \ell) B(m, n+\ell) = B(n + m, \ell), 
\ee
one can easily check that 
the Virasoro algebra is satisfied at the classical level (in Poisson brackets)
\be \label{QV}
(Q_m, Q_n) = - (m-n) Q_{m+n}. 
\ee

Note that the function $B(n, k)$ transforms under 
the similar transformation 
\be
Q_m \rightarrow U Q_m U^{-1}, \qquad 
U = e^{\sum_k \lam_k a_{-k} b_k}
\ee
as 
\be
B(n, k) \rightarrow \frac{e^{k\lam_k}}{e^{(n+k)\lam_{(n+k)}}}B(n, k).
\ee
Choosing 
\be \label{lamk}
\lam_k = \frac{1}{2k}\log[k/\dot{f}(ik)],
\ee
we get 
\be
B(n, k) \rightarrow 1.
\ee
This means that the function $f(\del_t)$ has no real physical significance,
except when (\ref{lamk}) is ill-defined, e.g. $\dot{f}(ik) = 0$ or $k = 0$. 

If $\dot{f}(ik) = 0$, 
$f$ has a degeneracy of zeros at $k$.
This is assumed to be not the case in our derivation.
In the above we have also assumed that $k \neq 0$ 
(see (\ref{omegaab})).
Hence, as far as we have considered, 
we can always set $B(n,k) = 1$ without loss of generality.

To summarize, 
for the action (\ref{Sxy}),
a nonlocal particle theory with conformal symmetry 
is characterized by its spectrum ${\cal Z}$.
All possible $f(\del_t)$'s with the same spectrum are equivalent
if $\dot{f}(ik)k > 0$ for all $k\in{\cal Z}$.
After change of variables,
the Virasoro algebra can always be realized 
in terms of the creation annihilation operators 
for excitation modes of this spectrum simply as
\be
Q_n = \sum_{k\in\mathbb{Z}} a_{-k} b_{k+n}.
\ee

Compared with the spectrum of an open string,
each sequence of zeros $\{ z_i + n: \; n \in \mathbb{Z} \}$
is expected to correspond to one dimension in the target space. 
The sequences come in pairs $(z_i, -z_i)$,
except when $z_i = 0$ or $z_i = 1/2$
(so that $\pm z_i$ define the same sequence).
In view of the spectrum of the sequences of zeros,
we expect that $z_i = 0$ corresponds to 
open strings with Neumann boundary conditions on both endpoints,
and $z_i = 1/2$ to open strings with Neumann boundary condition 
on one endpoint and Dirichlet boundary condition on the other endpoint.
This will be shown more explicitly below,
and we will discuss the generic case 
with $z_i \in (0, 1/2)$ in Sec. \ref{OpenString}.


Incidentally, we remark that the particle theory (\ref{Sxy}) does not 
exhaust all possible realization of the Virasoro algebra
in terms of creation and annihilation operators. 
We give a counter-example in the appendix.

The formulas above can be easily modified to adapt to 
the case of a single field
\be
x = y.
\ee
First, the function $f(\del_t)$ should be symmetric $f(\del_t) = f(-\del_t)$.
Thus the zeros of $f$ will come in pairs $(k, -k)$ 
\be \label{calZ}
{\cal Z} = \{ k \; | \; f(ik) = 0 \} = \{ \pm z_i + n; \, n \in \mathbb{Z} \},
\ee
except when the zeros are on the fixed points $\mathbb{Z}$ or $\mathbb{Z} + 1/2$. 
Without loss of generality we can assume that $z_i \in [0, 1/2]$.

The general solution of $x$ is 
\be
x = \frac{1}{\sqrt{2}} \sum_i \sum_{n\in\mathbb{Z}} \left( x_{(z_i + n)} e^{-i(z_i + n)t} + y_{(z_i + n)} e^{i(z_i + n)t} \right). 
\ee
If there are zeros on the fixed points, 
we get in the general solution of $x$ 
\be
x_0 + p t + \sum_{n \neq 0} x_n e^{int}, \qquad \mbox{and/or} \qquad 
\sum_{n \in \mathbb{Z}} x_{n+1/2} e^{i(n+1/2)t}. 
\ee

The simplectic two-form $\omega$ 
is given by the same expressions (\ref{omega4}) above.
The conserved charges become 
\be
Q_n = \sum_{k \in \cal Z} [(k+n)\dot{f}(ik) + k\dot{f}(i(k+n))] x_{-k} y_{k+n},
\ee
assuming that the fixed points do not appear in ${\cal Z}$.

\subsection{Quantization}

Let us consider Virasoro generators of the form
\be
L_n = \sum_{k \in \cal Z} B(n, k) :a_{-k} b_{k+n}:, 
\ee
where $a_k$, $b_k$ satisfy 
\be
[ a_k, b_{\ell} ] = k \d^0_{k+\ell}.
\ee
This is just (\ref{QB}) with normal ordering. 
For $k > 0$, $a_k, b_k$ are annihilation operators, 
and for $k < 0$, $a_k, b_k$ are creation operators. 
Then we have 
\be
[ L_m, a_k ] = - k B(m, -m-k) a_{m+k}, \qquad 
[ L_m, b_k ] = - k B(m, k) b_{m+k}. 
\ee
From this it is straightforward to check that if 
$B$ satisfies the identity 
\be
(k + m) B(m, k) B(n, k+m) - (k + n) B(n, k) B(m, k+n) 
= \frac{1}{2}(m - n) B(m + n, k),
\ee
the Virasoro algebra is guaranteed
\be
[ L_m, L_n ] = (m-n) L_{m+n} - \delta^0_{m+n} C(n), 
\ee
where 
\be
C(n) = - \sum_{-n<k<0, k\in\cal Z} k(k+n) B(n, k) B(-n, k+n). 
\ee
For $B(n, k) = 1$ , we get $C(n) = \frac{c}{12} n(n-1)(n+1)$ 
with the central charge $c = 2$ for each sequence of zeros of $f$, 
as it should be for two scalar fields $x$ and $y$.
For the generic expression of $B$ (\ref{B}) in the nonlocal particle theory, 
we have $B(n, k) B(-n, k+n) = 1$, 
and so 
\be
C(n) = \sum_{m=1}^n (m-z)(n-m+z) = \frac{1}{6} n(n-1)(n+1) + n z(1-z)
\ee
for the contribution of a sequence $k = z+m \; (m\in\mathbb{Z})$.
We can absorb the last term into $L_0$ 
by shifting $L_0$ by the constant $z(1-z)/2$.
The central charge is therefore always $c=2$ for each sequence of zeros. 

If there is only one scalar field, i.e., $x = y$, 
the contribution of each sequence of zeros to the central charge is $1$.

The contribution of a sequence of zeros of $f$ to the Casimir energy $a$
(ground state energy of the Hamiltonian) can be computed as
\be
a = \sum_{k} (a_{-k}b_k - :a_{-k}b_k:) 
= \sum_{k<0} (-k) = \sum_{m=1}^{\infty} (m-z) 
= \frac{1}{24} - \frac{1}{8}(2z-1)^{2}
= -\frac{1}{12} + \frac{1}{2} z(1-z).
\ee
When $z = 0$, this gives $a = -1/12$ as it should 
for 2 flat directions in the target space.

The $z$-dependent piece in $a$ suggests that 
we shift $L_0$ by the same number $z(1-z)/2$ 
as it was suggested by $C(n)$ above.
In terms of the shifted operator $L_0$, 
the contribution of a sequence of zeros of $f$ to
the central charge and the zero-point energy 
are the same as 2 flat directions in the target space.
Nevertheless,
the conformal field theory defined by a generic $f$
can certainly be inequivalent to 
an open string in a flat background,
because the spectrum $\{z_i + n\}$ and Virasoro constraints 
are different.

Given the same ghost system as the open string
(which will be described as ghost of a nonlocal particle below),
we should select 26 sequences of zeros of $f$ so that 
the total central charge of the Virasoro algebra is $0$.
The shifted $L_0$ should be defined as
\be
L_0 = \sum_i \left[ \sum_{k_i=z_i + \mathbb{Z}} :a_{-k_i}b_{k_i}: 
+ \frac{1}{2} z_i(1-z_i) \right],
\ee
and the Virasoro constraints should be
\be
L_m - a \delta_m^0 = 0, \qquad m \geq 0, 
\ee
where $a = \frac{S-T}{24}$.
Here $S$ and $T$ are the number of sequences of zeros of $f$ 
for target space coordinates which are space-like or time-like. 
Since $S + T = 26$ in order to cancel the Virasoro anomaly of the ghost, 
here we want $T = 1$ and $S = 25$, and so $a = 1$.

\subsection{Classification}

In the case $x = y$,
\be \label{S}
S = - \frac{1}{2\pi} \int dt \; x f(\del_t) x.
\ee
Without loss of generality we can assume that the function $f(\del_t)$ is even 
\be
f(-\del_t) = f(\del_t).
\ee
The equation of motion is 
\be \label{EOM}
f(\del_t) x = 0.
\ee
If the transformation (\ref{transf}) is a symmetry, 
given a solution $x$ of the equation of motion, 
$e^{int} \dot{x}$ must also be a solution.
That is, we need
\be \label{condition}
f(\del_t + in) \dot{x} = 0
\ee
whenever $x$ satisfies (\ref{EOM}). 

We will consider three classes of $f$ with conformal symmetry. 
These are the cases when the nonlocal particles 
would be naturally described as a string 
because, as we will see below, 
the symplectic two-form and conserved charges 
can be naturally expressed as integrals of a fictitious variable.
The mathematical identity that will help us to do this is
\be \label{identity}
A(t+a) B(t) - A(t) B(t-a)
= \frac{d}{dt}\left[ \int_0^a d\beta
A(t+\beta) B(t+\beta - a) \right].
\ee
To show how this formula leads to a fictitious dimension 
on the nonlocal particle, 
we will carry out the canonical quantization all over again, 
although the results above can be applied to all the examples 
of this section.

\subsubsection{First class}

For the first class, $f$ is a periodic or anti-periodic function
\be \label{class1}
f(\del_t + i) = \pm f(\del_t).
\ee
Given (\ref{class1}), the conformal symmetry condition 
(\ref{condition}) is equivalent to
\be
(\pm)^n \del_t[f(\del_t) x^{\mu}] = 0, 
\ee 
which holds whenever the equation of motion is satisfied. 
Since we have assumed that $f$ is an even function,
its Fourier expansion is
\be \label{first}
f(\del_t) = {\sum_n}' f_n e^{n\pi \del_t}, \qquad
f_{-n} = f_n,
\ee
where ${\sum_n}'$ denotes a sum over either even or odd numbers, 
depending on whether $f$ is periodic or anti-periodic.

Let us start with one space-time dimension for the first class. 
Variation of the action is
\begin{align}
\delta S &= \frac{1}{2\pi} \int dt \left\{ \delta x(t) f(\del_t) x(t) + x(t){\sum_n}' f_n \delta x(t+n\pi) \right\}\nn \\
&= \frac{1}{2\pi} \int dt \left\{ \delta x(t) f(\del_t) x(t) + {\sum_n}' f_n \left\{ \vphantom{\int}x(t-n\pi ) \delta x(t)
+\frac{d}{dt} \left[ \int^{n\pi}_0 d\beta \; x(t+\beta-n\pi)\delta x(t+\beta) \right] \right\}\vphantom{{\sum_{}}'}\right\} \nn\\
&= \frac{1}{2\pi} \int dt \left\{2 \delta x(t) f(\del_t) x(t) + 
\frac{d}{dt}{\sum_n}' f_n\left[ \int^{n\pi}_0 d\beta \; x(t+\beta-n\pi)\delta x(t+\beta)\right] \right\}.
\end{align}
The symplectic 2-form is the differential of the total derivative part of $- \d S$
\be 
\om = \frac{1}{2\pi} {\sum_n}' f_n \left[ \int_0^{n\pi} d\beta \; \d x(t+\b - n\pi) \d x(t+\b ) \right].
\ee
Noether's theorem gives us the conserved charges for
the symmetry transformations (\ref{transf})
\be \label{transfm} 
\delta t = \epsilon e^{-imt}
\ee
as
\begin{align}
Q_m &= \frac{1}{2\pi} {\sum_n}' f_n \left[ \int_0^{n\pi} d\b \; e^{-im(t+\b)} \dot{x}(t+\b) x(t+\b - n\pi) \right], 
\label{Qm1}
\end{align}
where the equation of motion is used to simply the expression.

\subsubsection{Second class}

The second class of actions with conformal symmetry 
is given by (\ref{S}) with 
\be \label{second}
f(\del_t) = g(\del_t) \del_t,
\ee
where $g(\del_t)$ is an odd periodic or anti-periodic function
\be \label{conditiong}
g(\del_t + i) = \pm g(\del_t), \qquad
g(-\del_t) = - g(\del_t).
\ee
The condition for conformal symmetry (\ref{condition}) is then 
\be
(\pm)^n (\del_t + in) [f(\del_t) x^{\mu}] = 0 
\ee
whenever $x$ satisfies the equation of motion. 
This is obviously valid. 
The Fourier expansion of $g$ is 
\be \label{Fourierg}
g(\del_t) = {\sum_{n\geq 0}}' g_n e^{n\pi \del_t}, \qquad
g_{-n} = -g_n.
\ee
Again the sum is only over either even or odd numbers.

Variation of the action is
\begin{align}
\delta S &= - \frac{1}{2\pi} \int dt \left\{ \delta x(t) f(\del_t) x(t) + x(t){\sum_n}' g_n \delta \dot{x}(t+n\pi) \right\}\nn \\
&= - \frac{1}{2\pi} \int dt \left\{ \delta x(t) f(\del_t) x(t) + {\sum_n}' g_n \left\{ \vphantom{\int}x(t-n\pi ) \delta \dot{x}(t)
+\frac{d}{dt} \left[ \int^{n\pi}_0 d\beta \; x(t+\beta-n\pi)\delta \dot{x}(t+\beta) \right] \right\}\vphantom{{\sum_{}}'}\right\} \nn\\
&= - \frac{1}{2\pi} \int dt \left\{2 \delta x(t) f(\del_t) x(t) + 
\frac{d}{dt}{\sum_n}'g_n\left[ -x(t+n\pi)\delta x(t) + \int^{n\pi}_0 d\beta \; x(t+\beta-n\pi)\delta \dot{x}(t+\beta)\right] \right\}.
\end{align}

The symplectic 2-form is thus 
\be \label{sym. two form}
\om = \frac{1}{2\pi} {\sum_n}' g_n \left[ -\d x(t+n\pi)\d x(t) + \int_0^{n\pi} d\beta \; \d x(t+\b - n\pi) \d \dot{x}(t+\b ) \right].
\ee
The conserved charges for the transformation (\ref{transfm}) 
are then
\begin{align}
Q_m &= \frac{1}{2\pi} {\sum_n}' g_n \left[ -x(t+n\pi) e^{- imt}\dot{x}(t) 
+ \int_0^{n\pi} d\b \; x(t+\b - n\pi) \del_\b (e^{- im(t+\b)}\dot{x}(t+\b)) \right] \nn \\	
&= \frac{1}{2\pi} {\sum_n}' g_n \left[-\int_0^{n\pi} d\b \; e^{- im(t+\b)} \dot{x}(t+\b) \dot{x}(t+\b - n\pi) \right.\nn \\
&\qquad \left. -(x(t-n\pi)+ x(t+n\pi)) e^{imt}\dot{x}(t)+ x(t)e^{- im(t+n\pi)} \dot{x}(t+n\pi)\vphantom{\int}\right] \nn \\
&= \frac{-1}{2\pi} {\sum_n}' g_n \left[ \int_0^{n\pi}  d\b \; e^{- im(t+\b)} \dot{x}(t+\b) \dot{x}(t+\b - n\pi) 
+ e^{- imt}(x(t-n\pi)+ x(t+n\pi))\dot{x}(t) \right] \nn\\
&= \frac{-1}{2\pi} {\sum_n}' g_n \left[ \int_0^{n\pi} d\b \; e^{- im(t+\b)} \dot{x}(t+\b) \dot{x}(t+\b - n\pi) \right]. \label{Qm}
\end{align}
Again the equation of motion is used to simply the expression. 

Since these are the conserved charges derived from
the symmetry (\ref{transfm}),
the Poisson brackets among the charges $Q_m$ are given
by the classical Virasoro algebra
\be
(Q_m, Q_n) = (m-n) Q_{m+n}.
\ee

Examples of this class of models include the nonlocal particle theory
equivalent to an open string with Neumann boundary conditions 
on both endpoints (see Sec. \ref{Neumann}),
or Neumann boundary condition on one endpoint 
but Dirichlet boundary condition on the other endpoint (\ref{Dirichlet}).

\subsubsection{Third class} \label{3rdclass}

The third class is the exceptional case mentioned above. 
It has 
\be
f(\del_t) = \frac{h(\del_t)}{\del_t}, 
\ee
where $h(\del_t)$ is an odd periodic function
\be
h(\del_t + i) = (\pm) h(\del_t), \qquad h(- \del_t) = - h(\del_t). 
\ee
The Fourier expansion of $h$ is the same as (\ref{Fourierg}). 
If $h$ has only a simple zero at $k \in \mathbb{Z}$, 
$f$ has no zero at $k = 0$, but has a zero at all other integers. 
Thus the sequence of zeros is not complete. 
Nevertheless,
the condition (\ref{condition}) is now 
\be
(\pm)^n (\del_t + in)^{-1} \del_t [h(\del_t) x(t)] = 0. 
\ee
This holds because the equation of motion $f(\del_t) x = 0$ implies that 
$h(\del_t) x = 0$. 

Since the inverse operator of $\del_t$ is not always well defined, 
we will only consider the cases when $h(0) = 0$, e.g. 
$h(\del_t) = \tanh(\pi\del_t)$. 
For the general solution
\be
x(t) = \sum_{n\neq 0} x_n e^{int},
\ee
it is straightforward to repeat the same derivation above to get
\be
\omega = C \sum_{n\neq 0} \frac{\d x_{-n} \d x_{n}}{in},
\ee
where $C = \dot{h}(i)$,
and 
\be
Q_m = C \sum_{n\neq 0} \frac{m+n}{n} x_{-n} x_{m+n}.
\ee
In this case it is not so easy to use the formula (\ref{identity})
to express $\om$ and $Q_m$ as an integral, 
because the inverse of $\del_t$ is not necessarily well defined.

An explicit example is the nonlocal particle equivalent to 
an open string with Dirichlet boundary conditions on both endpoints
(see comments at the end of Sec. \ref{Dirichlet}).

\subsection{More than one dimensions}

One can easily extend the nonlocal particle theory in one dimensions 
to $D$ dimensional spacetime, replacing (\ref{S}) by 
\be
S = - \frac{1}{2\pi} \int dt \; x^{\mu} f(\del_t) x_{\mu}.
\ee
This theory has Lorentz symmetry 
(but not necessarily translational symmetry).
If we do not care about Lorentz symmetry, 
there can be more general nonlocal action with conformal symmetry
\be
S = \frac{1}{2\pi} \int dt x^{\mu}f_{\mu\nu}(\del_t)x^{\nu},
\ee
where $f_{\mu\nu}$ satisfies
\be
f_{\mu\nu}(-\del_t) = f_{\nu\mu}(\del_t), \qquad \mu, \nu = 0, \cdots, d. 
\ee
The $N$-th class ($N = -1, 0, 1$) of action with conformal symmetry has 
\be
f_{\mu\nu}(\del_t) = g_{\mu\nu}(\del_t) \del_t^{N}, 
\ee 
where $g_{\mu\nu}(\del_t)$ satisfies 
\be
g_{\mu\nu}(\del_t + i) = \Lambda_{\mu}{}^{\lam} g_{\lam\nu}(\del_t), \qquad
g_{\mu\nu}(-\del_t) = (-1)^N g_{\nu\mu}(\del_t), 
\ee
and $\Lambda_{\mu}{}^{\lam}$ is an $SO(1,d)$ matrix.

\section{Unitarity} \label{Unitarity}

In the above we have only considered the mathematical formulation
of nonlocal particles, 
without worrying about whether they make physical sense or not.
As a simple example of how things can go wrong,
we consider the case with
\be
g(\del_t) = \sinh(\pi\del_t), \qquad
(g_1 = -g_{-1} = 1/2, \quad g_{n\neq 1} = 0,)
\ee
for an action of the 2nd class. 
The equation of motion is
\be
\dot{x}(t+\pi) = \dot{x}(t-\pi)
\ee
and the general solution is
\be \label{solution}
x(t) = x_0 + y_0 t + \sum_{n\neq 0}\frac{y_n}{in}e^{int}.
\ee
According to (\ref{sym. two form}), the symplectic 2-form is
\begin{align}
\om 
= \d x_0 \d y_0 + \sum_{n \neq 0} \frac{(-1)^n}{2in} \d y_{n} \d y_{-n}.
\end{align}

The inverse of the symplectic 2-form is the Poisson bracket
\be \label{Pyy}
(x_0,y_0)=1,\quad (y_n,y_{-n})=(-1)^{n}2in.
\ee
The Poisson bracket for $x$ is thus
\be \label{Pxx}
( x(t), \dot{x}(t'+\pi) ) = -1+\sum_{n\in\mathbb{Z}} 2 e^{in(t-t')} = -1 + 4\pi \, \delta(t-t').
\ee

Upon quantization, eq.(\ref{Pyy}) is equivalent to
\be \label{yy}
[y_m, y_n] = (-1)^m 2 m \delta_{m+n}^0.
\ee
The Virasoro generators are
\begin{align} \label{Qm1}
Q_m &=\frac{1}{4\pi}\left[-\int_0^{\pi}\dot{x}(t+\b-\pi)e^{- im(t+\b)}\dot{x}(t+\b)d\b
+ \int_0^{-\pi}\dot{x}(t+\b+\pi)e^{- im(t+\b)}\dot{x}(t+\b)d\b \right]\nn \\
&= \frac{-1}{4\pi}\left[ \int_{-\pi}^{\pi}(y_0+\sum_{n} y_n e^{in(t+\b-\pi)})e^{- im(t+\b)}(y_0+\sum_{n'} y_{n'} e^{in'(t+\b)})\right]\nn \\
&= \frac{-1}{2} \sum_n (-1)^n y_n y_{m-n}, 
\end{align}
which vanish for odd $m$. 
The Hamiltonian is
\be \label{Q0}
H = Q_0 = \frac{-1}{2} \sum_n (-1)^n y_{-n} y_n.
\ee
Apparently the operators $y_m$ and $Q_m$ are
reminiscent of $\a_m$ and $L_{-m}$ in the open string theory.
The only difference is the extra factors of $(-1)^n$'s, 
and this difference is crucial.
Due to these $(-1)^n$ factors,
either the Hilbert space contains negative-norm states,
or the Hamiltonian is unbounded from below. 
From (\ref{Q0}), we see that the Hamiltonian is unbounded from below 
due to oscillation modes with even $n$. 
One might want to avoid this pathology at the quantum level 
by defining the vacuum to be annihilated by $y_{-n}$ for even $n>0$, 
so that states created by $y_{n}$ for even $n$ has positive energy. 
For example, for $m>0$, 
the state $y_{2m} \left| 0 \right\rangle$ has positive energy 
\be
	H y_{2m}\left| 0 \right\rangle=-y_{2m}(-2m+y_{2m}y_{-2m})\left| 0 \right\rangle=2m y_{2m}\left| 0 \right\rangle. 
\ee
(The odd modes will then obey the usual convention that $y_{2m+1}| 0 \rangle = 0$ for $m \geq 0$.) 
But then this state is of negative-norm 
\be
	\left\langle 0 \right|y_{-2m} y_{2m} \left| 0 \right\rangle = -2m < 0. 
\ee

In fact, it is not necessary to go through the detailed computation to
tell whether an action suffers the problem of negative-norm states
or the problem of Hamiltonians unbounded from below.
For an ordinary free scalar theory in $D$ dimensions, 
the Fourier transformation of two-point correlator is give by
\begin{equation}
\langle \psi(p)\psi(q)\rangle =\delta^{(D)}(p-q)\frac{i}{p^2-m^2+i\varepsilon}. 
\end{equation}
Now if the propagator has multiple poles, 
the correlator near the $i$-th pole is given by
\begin{equation}
	\langle \psi(p)\psi(q)\rangle \simeq \delta^{(D)}(p-q)\frac{i C_i}{p^2-m_i^2+i\varepsilon}, 
\end{equation}
where $C_i$ is the residue of the $i$-th pole. 
The sign of this quantity when $p$ and $q$ are on-shell 
(which is the same as the sign of $C_i$ because $\varepsilon > 0$) is correlated with 
the sign of the norm of the single particle state with momentum $p$. 
Therefore the requirement of unitarity is equivalent to 
demanding the residues of all poles of the propagator, 
which is $1/f(ik)$ for our case, to be of the same sign. 
Otherwise, there will either be negative-norm states or the Hamiltonian will be unbounded from below. 

In the example above, $f(ik) = -k\sin(\pi k)$ and the residues of poles of the propagator $1/f(ik)$ 
have alternating signs. 
Hence half of the oscillation modes have negative energy while the other half have positive energy. 

In the following we will demand that the theory be unitary 
(no negative norm states)
and that the classical Hamiltonian be positive definite.
(Tachyons can appear after 
including the quantum correction due to Casimir effect, 
yet the quantum spectrum will still be bounded from below.)  
In other words, 
we will demand that the residues of poles of $1/f(ik)$ 
be all positive. 
When $k$ is close to a pole of $1/f(ik)$, we have 
\be
1/f(ik) \simeq \frac{1}{f'(ik_0)(k-k_0)}, 
\ee
and thus the residue of the pole at $k_0$ is just $1/f'(ik_0)$. 
This means that we shall demand that the slope of $f(ik)$ at all zeros of $f(ik)$ be positive.

\section{Open string as nonlocal particle} \label{OpenString}

\subsection{Neumann boundary condition} \label{Neumann}

The requirement of unitarity implies that $g(\del_t)$ must be periodic rather than anti-periodic, 
because the latter implies that half of the poles have negative residues. 
We also want $g(-\del_t) = -g(\del_t)$ (see (\ref{conditiong})) to hold.
A simple modification of our previous example that would meet our needs is 
\be \label{openstringg}
g(\del_t) = \tanh(\pi\del_t) = \frac{\sinh(\pi\del_t)}{\cosh(\pi\del_t)}.
\ee
The same action was considered by Kato \cite{Kato} long time ago 
as a particle action equivalent to the bosonic open string. 

Let us now derive this action from open string theory by 
integrating out the bulk degrees of freedom on the worldsheet, 
and show that this action is just the effective action for the boundary coordinates. 

For a quadratic action, the effect of integrating out a variable
is the same as plugging a classical solution into the action.
Thus, to integrate out $X(\tau, \s)$ in the bulk of the string,
we only need its solution to the equation of motion for given boundary value $X(\tau, 0) = X_0(\tau)$.
The general solution to the equation of motion
\be \label{worldsheetEOM}
(\del_{\tau}^2 - \del_{\s}^2) X(\tau, \s) = 0
\ee
with the Neumann boundary condition $\del_{\s} X = 0$ at $\s = \pi$ is
\be \label{Xs}
X(\tau, \s) = \int dk \; e^{ik\tau} cos(k(\s-\pi)) \tilde{x}(k),
\ee
where $\tilde{x}(k)$ is determined by $X_0$
\be \label{xk}
\tilde{x}(k) = \frac{1}{2\pi \cos(k\pi)} \int d\tau \; e^{-ik\tau} X_0(\tau).
\ee
Plugging (\ref{Xs}), (\ref{xk}) into the action, we get 
\bea
S &=& \frac{1}{2\pi} \int d\tau \; \int_0^{\pi} d\s \left( (\del^{\tau} X)^2 - (\del_{\s} X)^2 \right) 
= \frac{1}{2\pi} \int d\tau \; \left[ X \del_{\s} X \right]_{\s=0} \nn \\
&=& - \frac{1}{2\pi} \int d\tau \, d\tau' \;  X_0(\tau) G(\tau-\tau') X_0(\tau'),
\eea
where the kernal is
\be
G(\tau) = \frac{1}{2\pi} \int dk \; k \frac{\sin(k\pi)}{\cos(k\pi)} e^{ik\tau}.
\ee
It is easy to check that this is equivalent to 
the action (\ref{S}) with the function $f(\del_t) = g(\del_t) \del_t$ 
where $g(\del_t)$ is given by (\ref{openstringg}). 

Since open string vertex operators only live on the worldsheet boundary, 
in principle we can turn an open string theory in any open string background 
to a nonlocal particle theory. 

Note that not every different choice of $g(\del_t)$ defines a different theory.
The choice (\ref{openstringg}) is physically equivalent to any other choice of the form 
\be
g(\del_t) = g_0(\del_t) \sinh(\pi\del_t)
\ee
as long as $g_0$ satisfies the following criteria: 
\begin{itemize}
\item $g_0(ik)$ has no zeros (so that the solution to the equation of motion is still (\ref{solution})). 
\item $g_0(ik)$ is anti-periodic (so that $g(ik)$ is periodic). 
\item $g_0(ik) = g_0(-ik)$ (so that $g(ik)$ is an odd function).
\item $(-1)^n g_0(in \pi) > 0$ (so that $g'(ik)$ is always positive at all zeros of $g(ik)$, which are $k = n\pi$.)
\end{itemize}
These properties are all we need for the derivation below. 

The symplectic two-form (\ref{sym. two form}) is
\begin{align}
\omega 
&= C \left[2 \d x_0 \d y_0 + \sum_{n\neq 0} \frac{1}{in} \d y_n \d y_{-n} \right], 
\end{align}
where 
\be
C \equiv {\sum_n}' \frac{n}{2} g_n = \frac{1}{2\pi} g'(0).
\ee
Canonical quantization therefore gives
\be
[ y_m, y_n ] = m C^{-1} \delta_{m+n}^0, \qquad [x_0,y_0]=\frac{1}{iC}.
\ee
According to (\ref{Qm}), the Virasoro generators are
\begin{align}
Q_m 
&= C \sum_{n \in\mathbb{Z}} y_{-m-n} y_n. 
\end{align}
Up to overall constant factors of $C$ which can be absorbed by scaling the coordinate $x$, 
these equations are the same as the open string with the following identification 
(assuming that we scaled $C$ to $C = 1$)
\be
x_0 = x, \qquad y_0 = p, \qquad y_m = \alpha_m, \qquad Q_m = 2 L_{-m}.
\ee 

\subsection{Dirichlet boundary condition} \label{Dirichlet}

For an open string with one endpoint (at $\s = \pi$) obeying Dirichlet boundary condition, 
we can also integrate out the bulk of the string to obtain 
the effective action for the other endpoint coordinate at $\s = 0$ 
\be
X(\t, \pi) = 0, \qquad X(\t, 0) = X_0(\t).
\ee
The general solution of the equation of motion (\ref{worldsheetEOM}) 
with the Dirichlet boundary condition is 
\be
X(\tau, \s) = \int dk \; e^{ik\t} \sin(k(\pi-\s)) \tilde{x}(k). 
\ee
Therefore, 
\be
\tilde{x}(k) = \frac{1}{2\pi\sin(k\pi)} \int d\t \; e^{-ik\t} X_0(\t). 
\ee
and we can plug it into the worldsheet action to get 
\be
S = - \frac{1}{2\pi} \int d\t d\t' X_0(\t) G(\t-\t') X_0(\t'), 
\ee
where 
\be
G(\t) = \frac{1}{2\pi} \int dk \; k \frac{\cos(k\pi)}{\sin(k\pi)} e^{ik(\t)}. 
\ee
This corresponds to 
\be
g(\del_t) = \frac{\cosh(\pi\del_t)}{\sinh(\pi\del_t)}
\ee
for our particle action (\ref{S}) with $f(\del_t) = g(\del_t) \del_t$.  

The general solution to the equation of motion for the nonlocal particle is 
\be
x(t) = \sum_{n \in \mathbb{Z}} x_{n+1/2} e^{i(n+1/2)t}. 
\ee
It follows that the symplectic two-form is 
\be
\om = \frac{1}{2} \sum_{n\in\mathbb{Z}} \frac{\d y_{n+1/2} \d y_{-n-1/2}}{n+1/2}, 
\ee
and the conserved charges are 
\be
Q_m = \frac{1}{2} \sum_{n\in\mathbb{Z}} \d y_{n+1/2} \d y_{m-n-1/2}. 
\ee

One might wonder what happens to the open string 
with Dirichlet boundary conditions on both endpoints?
In this case we cannot apply the same trick used above.
But we can construct the equivalent nonlocal particle theory directly 
from the spectrum of the open string, 
which is $\mathbb{Z}-\{0\}$.
This spectrum suggests a model of the third class
with $h(\del_t) = \tanh(\pi\del_t)$.
It is obvious that this nonlocal particle theory 
is equivalent to the open string theory 
with Dirichlet boundary conditions on both endpoints.

\subsection{Ghosts} \label{Ghosts}

To fully match with the open string theory,
we also need to rewrite the ghosts living on the string worldsheet as a particle theory.
The action for the 2 dimensional $bc$ conformal field theory should be
\be \label{ghostaction}
S_g = \frac{i}{2\pi} \int dt \; b(t) g(\del_t) c(t),
\ee
where $b$ and $c$ are anticommuting fields.
The action is invariant under the symmetry transformation
\be
\d b(t) = \epsilon e^{int} \big(\dot{b}(t) + in\lambda b(t)\big), \qquad
\d c(t) = \epsilon e^{int} \big(\dot{c}(t) + in(1-\lambda) c(t)\big),
\ee
where $(\lambda, 1-\lambda)$ are the conformal weights of $b$ and $c$.

It is straightforward to derive the symplectic two-form
\be \label{ombc}
\om = \frac{i}{2\pi} {\sum_n}' g_n \int_0^{n\pi} d\beta \;
\d b(t+\b-n\pi) \d c(t+\b),
\ee
and the conserved charges
\be \label{Qbc}
Q_{m} = \frac{i}{2\pi} {\sum_n}' g_n \int_0^{n\pi} d\b\; e^{im(t+\b)}
b(t+\b-n\pi) \left[ \dot{c}(t+\b) + im (1-\lambda) c(t+\b) \right].
\ee
Substituting the general solutions of the equation of motion
\be
b=\sum_{n=-\infty}^{\infty}b_n e^{-int},\quad c=\sum_{n=-\infty}^{\infty}c_n e^{-int},
\ee
into the symplectic two-form (\ref{ombc}) and conserved charges (\ref{Qbc}), one finds 
\begin{align}
  \om &= iC \sum_{n=-\infty}^{\infty}b_n c_{-n}, \\
  Q_{m}&= C \sum_{n=-\infty}^{\infty} [m(\lambda-1)-n] b_{m+n}c_{n}. 
\end{align}
We should take $\lambda = 2$ for the ghosts of open string theory.
The 26 dimensional bononic open string theory
can thus be viewed as a 26 dimensional theory of nonlocal particles
with the action (\ref{S}) of the second class (\ref{second})
with $g(\del_t)$ given by (\ref{openstringg}),
plus a ghost with the action (\ref{ghostaction}).

Now that we have all the algebraic elements of open string theory, 
we can construct vertex operators and define scattering amplitudes 
in exactly the same way we define perturbative open string theory. 
The vertex operator is defined as a local operator $V$ with conformal dimension 1
\begin{equation}
	[Q_m,V]=e^{imt}\left(-i\frac{d}{dt}+m\right)V.
\end{equation}
For example, the tachyon vertex should be given by
\begin{align} 
	V_0(k,t)&=:e^{ik\cdot  X(t)}:\nn\\
	&=\exp\left(k\cdot \sum_{n=1}^\infty \frac{y_{-n}}{n} e^{int}\right) \, e^{ik \cdot x(t)} \, \exp\left(-k\cdot \sum_{n=1}^\infty \frac{y_{-n}}{n}e^{-int}\right) ,
\end{align}
where
\begin{equation}
	k^2=2,\qquad X^\mu(t) = {x}_0^{\mu} + y_0^\mu t + \sum_{n\neq 0}\frac{y_n^\mu}{in}e^{int},
	\qquad x^\mu(t)={x}_0^{\mu}+y_0^\mu t. 
\end{equation}
Obviously all the scattering amplitudes 
are exactly the same as in open string theory at the tree level. 
The path integral approach to loop diagrams based on Riemann surfaces of different topology is obscured. 
However we can still take the operator approach \cite{GSW} to construct the loop amplitudes. 
Anyway, there is a unique way to define loop diagrams without breaking conformal symmetry. 

Fermionic fields can also be introduced on the worldline 
in a similar way as the ghosts. 
It should be straightforward to define nonlocal particles
equivalent to superstrings.
The worldsheet theory of a superstring 
is equivalent to the juxtaposition of free bosons, 
free fermions and ghosts, 
all of which we know how to describe 
in terms of nonlocal particles. 
Realizing that the oscillation modes on a string 
can be directly matched with those on a nonlocal particle, 
to derive the supersymmetry transformation rule
for the nonlocal particle theory, 
one can simply copy the transformation rules 
for the oscillation modes on a superstring, 
and reinterpret them as the rules for the oscillation modes 
on a nonlocal particle.

Before closing this section, 
we also comment that, 
while the notion of a 2 dimensional metric is absent 
in the particle theory, 
it seems impossible to talk about anything
more than what is in the conformal gauge, 
e.g. the Weyl symmetry.
On the other hand, since we have introduced 
a fictitious coordinate $\beta$ 
in expressions like (\ref{Qm}), 
and thus the 2 dimensional nature of the nonlocal particle 
is not totally invisible, 
it may be possible to introduce auxiliary fields 
to ``covariantize'' these expressions, 
such that these auxiliary fields correspond to 
the 2 dimensional metric on a string before gauge fixing. 
At that time we can start 
addressing questions concerning Weyl symmetry etc.

\section{Open string in nontrivial backgrounds} \label{ThirdExample}

In the above we have shown that a certain choice of the function $f(\del_t)$
in the nonlocal particle action (\ref{S}) leads to a theory equivalent  
to the bosonic open string in flat space with the Neumann (or Dirichlet) boundary condition.
A different choice of $f(\del_t)$ would then lead to a theory equivalent to  
the bosonic open string in flat space with a nontrivial boundary interaction, 
corresponding to a nontrivial open string background, i.e.,
a nontrivial D-brane configuration.
We will continue to focus on quadratic actions only.

Let us consider an explicit example where the action is of second class with 
\begin{equation} \label{gnontrivial}
	g(\del_t)=\frac{\sinh(\pi \del_t+i\theta)\sinh(\pi \del_t-i\theta)}{\sinh(2\pi \del_t)}. 
\end{equation}
The general solution to the equation of motion is 
\be \label{twistedx}
x(t)=\sum_{n \in \mathbb{Z}} \frac{y_n^{+}}{i(n-\theta/\pi)}e^{i(n-\theta/\pi)t} + \sum_{n \in \mathbb{Z}} \frac{y_n^{-}}{i(n+\theta/\pi)}e^{i(n+\theta/\pi)t}.
\ee
The symplectic two-form is
\begin{align}
\om 
= \sum_{n \in \mathbb{Z}} \frac{\d y^+_n \d y^-_{-n}}{i(n - \th/\pi)}.
\end{align}
The conserved charges are 
\be
Q_l = 
- \sum_{n \in \mathbb{Z}} y_n^{+} y_{-n-l}^{-}.
\ee
The expression of the symplectic two-form suggests that we decompose $x(t)$ in another way
\begin{equation}
	x(t) = \tilde{x}^+(t) + \tilde{x}^-(t),
\end{equation}
where
\begin{equation}
	\tilde{x}^{\pm} = \sum_n \frac{\tilde{y}^{\pm}_n}{i(n\mp\sgn(n)\theta/\pi)} e^{i(n\mp\sgn(n)\theta/\pi)t},
\end{equation}
where the function $\sgn(n)$ is defined by $\sgn(n) \equiv n/|n|$ for $n\neq 0$ and $\sgn(0) = 1$, and
\begin{equation}
	\tilde{y}^{\pm}_n = y^{\pm}_n \qquad (n \geq 0), \qquad
	\tilde{y}^{\pm}_n = y^{\mp}_n \qquad (n < 0).
\end{equation}
Upon quantization, we get
\be
[ \tilde{y}^+_0, \tilde{y}^-_0 ] = -\frac{\th}{\pi}, 
\ee
and
\begin{equation} 
	[\tilde{y}^+_n,\tilde{y}^+_{-n}] = n-\theta/\pi, \qquad
	[\tilde{y}^-_n,\tilde{y}^-_{-n}] = n+\theta/\pi \qquad \mbox{for} \quad n > 0. 
\end{equation}

Let us make a digression here and consider the limit $\theta \rightarrow 0$ 
and change of variables 
\begin{equation}
	x_0\equiv \frac{\tilde{y}^+_0-\tilde{y}^-_0}{-2i\theta/\pi},\qquad y'_0\equiv \tilde{y}^+_0+\tilde{y}^-_0. 
\end{equation}
Commutation relations are now 
\begin{align}
	[x_0,y'_0] 
	=-\frac{i}{2C(\theta)}, \qquad
    [\tilde{y}^+_n,\tilde{y}^+_{-n}]=\frac{n}{2C(\theta)}, \qquad
	[\tilde{y}^-_n,\tilde{y}^-_{-n}]=\frac{n}{2C(\theta)}.
\end{align}
Interestingly, the operators have a one-to-one correspondence with those of the closed string 
\begin{align}
x_0 = x, \qquad  y'_0 = p, \qquad
\tilde{y}^+_m = \a_m, \qquad \tilde{y}^-_m = \tilde{\a}_m, 
\end{align}
where the notation is such that the closed string target space variable is 
\be
X(\t, \s) = x + p \t + \frac{i}{\sqrt{2}} \sum_{n\neq 0} \left[ \frac{\a_m}{m} e^{im(\t+\s)} + \frac{\tilde{\a}_m}{m} e^{im(\t-\s)} \right].
\ee
It appears that the closed string may be viewed as 
a special limit of a more general nonlocal particle model. 
However this observation has to be taken with a grain of salt, 
because the closed string theory 
has two commuting copies of Virasoro algebra for 
the left-moving and right-moving modes, 
but there is only one copy of Virasoro algebra for this particle theory.
We will see how to generalize our ansatz for nonlocal particles 
to incorporate closed strings.

This nonlocal particle theory is actually equivalent to an open string theory
with the usual bulk Lagrangian in the flat space 
and a nonlocal boundary interaction 
\be
S = \frac{1}{2\pi} \int d\tau d\sigma \del X \bar{\del} X 
+ \left. \int d\tau X h(\del_\t) \dot{X} \right|_{\sigma=0}, 
\ee
where $h(\del_\t)$ is the difference between (\ref{gnontrivial}) and (\ref{openstringg})
\be
h(\del_\t) = \frac{\sinh(\pi \del_t+i\theta)\sinh(\pi \del_t-i\theta)}{\sinh(2\pi \del_t)}
- \frac{\sinh(\pi\del_t)}{\cosh(\pi\del_t)}.
\ee
This should be obvious from our derivation in Sec. \ref{Neumann}.
We have thus found an exact, 
marginal (nonlocal) deformation of the bosonic open string theory.
Nonlocal interactions in string theory have been studied 
in the past in different contexts \cite{NonlocalString}.

However, in Sec. \ref{Generators} we commented that 
the only thing that matters physically in the quadratic nonlocal particle action 
is the spectrum.
The description above is not giving the essential properties of the theory
in the most transparent way as it depends on too much detail of $f$.
Instead, we can try to find a string worldsheet theory with the same spectrum.
Given the two sequences of zeros $\{n \pm \th/\pi\}$,
it is natural to construct a worldsheet scalar field 
\be
Z(\t, \s) = \sum_{n\in\mathbb{Z}} 
\left[ \a_n e^{i(n+\th/\pi)(\t+\s)}
+ \b_n e^{-i(n+\th/\pi)(\t-\s)} \right], 
\ee
which gives the same general solution as (\ref{twistedx})
\be
Z_0(\t) = \sum_{n\in\mathbb{Z}} 
\left[ \a_n e^{i(n+\th/\pi)\t}
+ \b_n e^{-i(n+\th/\pi)\t} \right]
\ee
at $\s = 0$, 
and it satisfies the twisted periodic boundary condition
\be
Z(\t, \s+2\pi) = e^{i2\th} Z(\t, \s).
\ee
This is a boundary condition consistent with 
the standard worldsheet action in flat space, 
that is, it satisfies the condition
\be
\left.( \d Z \del_{\s} Z^* + \d Z^* \del_{\s} Z )\right|_{\s = 0}^{\s = 2\pi} = 0
\ee
necessary to guarantee that a solution of the equation of motion
\be
\del \bar{\del} Z = 0
\ee
extremizes the action.

Note that the twisted periodic boundary condition 
is not a typical open string boundary condition, 
which should not have any nonlocal correlation between 
the two endpoints of the string.
This is, for example, not the same as the boundary conditions 
for an open string stretched between two D-branes 
separated by an angle $\th$ in the complex plane of $Z$, 
because the values of $Z$ at $\s = 0, 2\pi$ are not fixed.
This is certainly not a closed string, either, 
since the string is not closed.
This is a generalization of both open string and closed string, 
where the distinction between them is blurred.

\section{Closed string as nonlocal particle} \label{ClosedString}

For the open string, 
we can interpret the nonlocal particle as the boundary of the open string. 
The equivalence between open string theory and particle theory is thus understandable. 
However, for the closed string, 
vertex operators can be inserted at any point on the worldsheet. 
No point on the closed string is special. 
It is hard to imagine how the closed string can also be equivalent to a nonlocal particle. 
Remarkably, 
we will show here that closed strings 
can also be described as nonlocal particles. 

Roughly speaking, the degrees of freedom and its symmetry in a closed string is 
the same as two copies of those of an open string, 
but with the zero modes of the two open strings identified. 
Thus it is natural to consider the particle action 
\be \label{closedstringS}
S = \frac{1}{2\pi} \int dt \; \left[ x(t) f(\del_t) x(t) + y(t) f(\del_t) y(t) 
+ \lambda \left(\sinh(\pi\del_t)(x(t) - y(t))\right) \right], 
\ee
where 
\be
f(\del_t) = g(\del_t) \del_t, \qquad g(\del_t) = \tanh(\pi\del_t). 
\ee
If the last term is absent in the action, 
we just have two copies of open string (with Neumann boundary condition). 
We shall identify them with the left-moving and right-moving modes 
on the closed string. 
There are also two copies of the conformal symmetry and its Virasoro algebra generators 
for $x$ and $y$ independently. 
The last term in (\ref{closedstringS}) involves a Lagrange multiplier $\lambda$. 
Note that since $\lambda$ is a constant variable, 
the last term is a total derivative. 
Hence the equations of motion for $x$ and $y$, 
as well as the symplectic two-form and conserved charge $Q_m$ are not modified. 
But it imposes the constraint 
\be
\int dt \; \left(\sinh(\pi\del_t)(x(t) - y(t))\right) = 0. 
\ee
Plugging in solutions of $x$ and $y$ 
\be
x = x_0 + p_0 t + \sum_n x_n e^{int}, \qquad y = y_0 + q_0  t+ \sum_n y_n e^{int}, 
\ee
the constraint implies that 
\be
p_0 = q_0. 
\ee
According to the symplectic two-form
\be
\om = x_0 p_0 + y_0 q_0 + \cdots = (x_0 + y_0) p_0 + \cdots,
\ee 
this implies that the conjugate variable of $p_0$ is now $(x_0+y_0)$. 
The other linear combination $(x_0-y_0)$ is a non-dynamical constant 
which can be dismissed. 
As we have achieved the identification of the zero modes, 
the closed string is thus also described as a nonlocal particle theory.

\section{Discussion} \label{Conclusions}

\subsection{Summary of nonlocal particle theory}

All physical properties of 
a nonlocal particle theory for two fields $x$ and $y$
with conformal symmetry are encoded in 
the spectrum, i.e.,
the zeros of $f$
\be
{\cal Z} = \{ z_i + n: \; n \in \mathbb{Z} \},
\ee
with each sequence specified by a number $z_i \in [0, 1)$. 
Here we assume that $\dot{f}(ik)k$ is of the same sign at all zeros of $f$,
otherwise the theory is not unitary.

If $x = y$, $f$ is symmetric and a zero at $k$ implies a zero at $(-k)$. 
The data characterizing the conformal theory of a single scalar field $x$ is 
again just
the zeros of $f$
\be
{\cal Z} = \{ z_i + m, \, - z_i - m : \; m \in \mathbb{Z} | f(z_i) = 0 \}
\ee
specified by a set of numbers $z_i$.
Each $z_i$ specifies two sequences of zeros of $f$ unless $z_i = 0$ or $z_i = 1/2$.
Other properties of $f$ are irrelevant. 
If there are two functions $f_1(\del_t)$ and $f_2(\del_t)$ 
with the same set of zeros and they differ from each other 
only by a function $\Delta f$ which 
vanishes at all the zeros of $f_1$ and $f_2$, 
the two theories defined by $f_1$ and $f_2$ are physically equivalent.
For example, if the zeros of $f$ are at $\{z+n, \; n \in \mathbb{Z}\}$,
then
\be
f(ik) + a \sin^2(\pi(k-z))
\ee
is equivalent to $f(ik)$.

\subsection{Nontrivial string backgrounds}

As it is easy to write down nonlocal particle actions 
which are equivalent to an open string in flat space with 
nonlocal interactions on the boundary
(i.e., those with a spectrum of $z \neq 0$ or $1/2$),
it will be very interesting to characterize the D-brane configurations 
corresponding to these nonlocal boundary interactions.
It will also be interesting to study Witten's cubic string field theory \cite{cubic}
for these open string backgrounds.

We have demonstrated that by considering particle theory with higher derivatives, 
it is possible to obtain not only open but also closed bosonic string theory. 
Using this method, one can easily find new consistent backgrounds of string theory.
Since solvable string backgrounds used to be rare, 
it will be very helpful to investigate string theory in these backgrounds 
for our understanding on the issue of
background (in)dependence of string theory.

\subsection{Boundary string field theory}

From the viewpoint of boundary string field theory,
we are considering a class of solvable backgrounds 
with nonlocal interactions on the boundary. 
This kind of backgrounds was considered by Li and Witten \cite{LiWitten}.
It will be interesting to see if the nonlocal particle theories 
with conformal symmetry
are solutions minimizing the boundary string field theory action.

In the boundary string field theory, 
the coefficients in the boundary interactions 
(coefficients in the function $f(\del_t)$)
correspond to spacetime fields.
The fact that apart from the location of zeros, 
most details of $f$ are physically irrelevant indicates 
a large gauge symmetry in the boundary string field theory.

\subsection{Compactification and other backgrounds}

While we claimed that we have constructed a closed string theory 
out of nonlocal particles,
an immediate question is whether we can consider 
compactification of space and find winding modes in the particle theory. 
This is a question that we will focus on in the near future. 
For the time being let us comment that the winding mode degrees of freedom 
can be easily found at the self-dual radius. 
If we compactify the target spaces of both $x$ and $y$ at the self-dual radius, 
both $p_0$ and $q_0$ are quantized. 
The linear combination $p_0+q_0$ corresponds to the momentum mode
and $p_0-q_0$ to the winding mode. 
Earlier we imposed the constraint $p_0 = q_0$ by a Lagrange multiplier $\lam$. 
We can just drop the Lagrange multiplier to recover the winding modes 
in the particle theory. 
Once we can construct the string field theory at the self-dual radius, 
it should be possible to find configurations corresponding to
compactifications at other values of the radius
via a study of the associated Higgs mechanism. 

On the other hand, it might not be possible to present 
all possible backgrounds of string theory as a nonlocal particle theory. 
The conformal field theory in the appendix with Virasoro generators 
defined by (\ref{QB}) 
with $B$ defined by (\ref{Buv}) is such an example. 
The value of the nonlocal particle presentation of the string theory 
lies in its simplicity and abundance of new backgrounds.

\subsection{Relation to rigid strings}

There was an attempt \cite{rigidstring} to describe the QCD string 
by the so-called ``rigid string'' which has a worldsheet action
involving higher derivative interactions.
However there are only finite derivatives, 
and its high temperature partition function was shown 
to have a wrong sign \cite{PolchinskiYang}.
A natural question is whether it is possible to construct 
actions with infinite derivative (nonlocal) terms  
to reproduce the correct high energy limit 
of a QCD flux tube \cite{Polchinski}, 
and whether such strings can be described 
as a nonlocal particle. 

Furthermore, while the rigid string can be viewed as 
a truncated description of a membrane with one direction 
wrapped on a compactified dimension \cite{Lindstrom}, 
later it was shown that a ``rigid string'' 
can also be viewed as derived from a string 
in a similar fashion \cite{Pavsic}.
The nonlocal string which is precisely equivalent to a string 
can thus be viewed as a completion of the approximation 
of a string by a ``rigid particle'', 
which has only finite higher derivative terms. 

Another topic that we will leave for future study 
is to find a nonlocal string/particle theory
which is equivalent to a membrane. 
Since the quantum membrane theory is much less 
understood than strings, 
this equivalence may provide useful help 
for us to better understand the M theory.

\section*{Acknowledgment}

The author thanks Chuan-Tsung Chan,
Kazuyuki Furuuchi, 
Hiroyuki Hata, Takeo Inami, Nobuyuki Ishibashi,
James Liu, Satchitananda Naik,
Shunsuke Teraguchi, 
Wen-Yu Wen and Syoji Zeze for helpful discussions.
This work is supported in part by
the National Science Council,
and the National Center for Theoretical Sciences
(NSC 94-2119-M-002-001), Taiwan, R.O.C.
and the Center for Theoretical Physics
at National Taiwan University.

\appendix

\section{A non-particle example of conformal symmetry} \label{Bexample}

For generators of the form (\ref{QB}), 
without assuming (\ref{B}), 
the Virasoro algebra (\ref{QV}) is satisfied if 
\be \label{Beq}
(k + m) B(m, k) B(n, k+m) - (k + n) B(n, k) B(m, k+n) 
= \frac{1}{2}(m - n) B(m + n, k). 
\ee
This condition has a nontrivial solution
\be \label{Buv}
B(n, k) = 1 + \frac{u + v n}{k} 
\ee
for constant parameters $u, v$. 
This can never be realized in a nonlocal particle theory 
because it does not satisfy (\ref{BBB}).
In order for $Q_0$ to have a non-negative spectrum, 
we would like 
\be
B(0, k) = 1+ \frac{u}{k}
\ee
to be positive-definite for all $k\in{\cal Z}$ (\ref{calZ}).
For $0 < z_i <1$, this is possible if 
\be
0< u < 1-z_i, \qquad \mbox{or} \qquad 
-z_i < u < 0. 
\ee

Upon quantization, the conserved charges are
\be \label{LB}
L_n = \sum_{k\in\cal Z} B(n, k) \, :a_{-k} b_{k+n}:. 
\ee
the Virasoro algebra is
\be
[ L_m, L_n ] = (m-n) L_{m+n} - \delta^0_{m+n} C(n), 
\ee
where 
\be
C(n) = - \sum_{-n<k<0, k\in\cal Z} k(k+n) B(n, k) B(-n, k+n) 
= \frac{1}{6}n(n+1)(n-1) + n(z+u)(1-z-u). 
\ee
This corresponds to $c=2$, 
and suggests a shift of $L_0$ by 
$(z+u)(1-z-u)/2$.
The central charge is $1$.


\vskip .8cm
\baselineskip 22pt
\begin{thebibliography}{99}
\itemsep 0pt

\bibitem{HoTian}
  P.~M.~Ho and Y.~Y.~Tian,
  ``UV-finite scalar field theory with unitarity,''
  JHEP {\bf 0501}, 026 (2005)
  [arXiv:hep-th/0410248].

\bibitem{Ho}
  P.~M.~Ho,
  ``Virasoro algebra for particles with higher derivative interactions,''
  Phys.\ Lett.\ B {\bf 558}, 238 (2003)
  [arXiv:hep-th/0208182].

\bibitem{EW}
  D.~A.~Eliezer and R.~P.~Woodard,
  ``The Problem of Nonlocality in String Theory,''
  Nucl.\ Phys.\  B {\bf 325}, 389 (1989).

\bibitem{ChengHoYeh}
  T.~C.~Cheng, P.~M.~Ho and M.~C.~Yeh,
  ``Perturbative approach to higher derivative and nonlocal theories,''
  Nucl.\ Phys.\  B {\bf 625}, 151 (2002)
  [arXiv:hep-th/0111160];
  T.~C.~Cheng, P.~M.~Ho and M.~C.~Yeh,
  ``Perturbative approach to higher derivative theories with fermions,''
  Phys.\ Rev.\  D {\bf 66}, 085015 (2002)
  [arXiv:hep-th/0206077].

\bibitem{Kato}
  M.~Kato,
  ``Particle Theories With Minimum Observable Length And Open String Theory,''
  Phys.\ Lett.\ B {\bf 245}, 43 (1990).

\bibitem{GSW}
  M.~B.~Green,J.~H.~Schwarz and E.~Witten,
  ``Superstring Theory,''
  (Cambridge Univ. Press, Cambridge, U.K.), Vols. I
  
\bibitem{NonlocalString}
  O.~Aharony, M.~Berkooz and E.~Silverstein,
  ``Multiple-trace operators and non-local string theories,''
  JHEP {\bf 0108}, 006 (2001)
  [arXiv:hep-th/0105309];
  O.~Aharony, M.~Berkooz and E.~Silverstein,
  ``Non-local string theories on AdS(3) x S**3 
  and stable non-supersymmetric backgrounds,''
  Phys.\ Rev.\  D {\bf 65}, 106007 (2002)
  [arXiv:hep-th/0112178];
  J.~L.~F.~Barbon and C.~Hoyos,
  ``AdS/CFT, multitrace deformations 
  and new instabilities of nonlocal string theories,''
  JHEP {\bf 0401}, 049 (2004)
  [arXiv:hep-th/0311274].

\bibitem{cubic}
  E.~Witten,
  ``Noncommutative Geometry And String Field Theory,''
  Nucl.\ Phys.\  B {\bf 268}, 253 (1986).

\bibitem{LiWitten}
  K.~Li and E.~Witten,
  ``Role of short distance behavior in off-shell open string field theory,''
  Phys.\ Rev.\  D {\bf 48}, 853 (1993)
  [arXiv:hep-th/9303067].

\bibitem{rigidstring}
  A.~M.~Polyakov,
  ``Fine Structure of Strings,''
  Nucl.\ Phys.\  B {\bf 268}, 406 (1986); 
  H.~Kleinert,
  ``The Membrane Properties of Condensing Strings,''
  Phys.\ Lett.\  B {\bf 174}, 335 (1986).

\bibitem{PolchinskiYang}
  J.~Polchinski and Z.~Yang,
  ``High temperature partition function of the rigid string,''
  Phys.\ Rev.\  D {\bf 46}, 3667 (1992)
  [arXiv:hep-th/9205043].

\bibitem{Polchinski}
  J.~Polchinski,
  ``High Temperature Limit Of The Confining Phase,''
  Phys.\ Rev.\ Lett.\  {\bf 68}, 1267 (1992)
  [arXiv:hep-th/9109007].
  
\bibitem{Lindstrom}
  U.~Lindstrom,
  ``DERIVING POLYAKOV'S RIGID STRING FROM A MEMBRANE,''
  Phys.\ Lett.\  B {\bf 218}, 315 (1989).

\bibitem{Pavsic}
  M.~Pavsic,
  ``On The Consistent Derivation Of Rigid Particles From Strings,''
  Class.\ Quant.\ Grav.\  {\bf 7}, L187 (1990).


  
\end{thebibliography}
\end{document}